\documentclass{article}              

\usepackage{graphicx}
\usepackage[bottom]{footmisc} 
\usepackage{natbib}
\usepackage{setspace}
\usepackage{fullpage}
\usepackage{bm}
\usepackage{amsmath} 
\usepackage{amssymb}
\usepackage{graphicx}
\usepackage{hyperref}
\usepackage{color}
\usepackage{float} 
\usepackage[labelsep=period]{caption}
\usepackage{url} 
\usepackage{rotating}
\usepackage{slashbox} 
\usepackage{caption} 
\usepackage{multicol} 
\usepackage{authblk}
\usepackage{graphics}

\newcommand{\re}{\mathbb R}
\newtheorem{algorithm}{Algorithm}
\newtheorem{thm}{Theorem}[section]
\newtheorem{prop}[thm]{Proposition}

\newtheorem{defn}[thm]{Definition}
\newtheorem{rem}[thm]{Remark}
\begin{document}

\title{Unsupervised classification of children's bodies using currents}

\author{
Sonia Barahona$^{(1)}$, Ximo Gual-Arnau$^{(2)}$,  Maria Victoria Ib\'a\~nez $^{(3)}$ and Amelia Sim\'o$^{(3)}$ \\

        \small{(1)     Department of Mathematics. Universitat Jaume I.
              Avda. del Riu Sec s/n. 12071-Castell\'on, Spain.}
                           \\
          \small{(2)   Department of Mathematics-INIT. Universitat Jaume I.
              Avda. del Riu Sec s/n. 12071-Castell\'on, Spain.}
                                      \\
         \small{(3)    Department of Mathematics-IMAC. Universitat Jaume I.
              Avda. del Riu Sec s/n. 12071-Castell\'on, Spain.}

    }

\maketitle

\begin{abstract}
Object  classification according to their shape and size is of key importance in many scientific fields. This work focuses on the case where the size and shape of an object is characterized by a \emph{current}. A \emph{current} is a mathematical object which has been proved relevant to the modeling of geometrical data, like submanifolds, through integration of vector fields along them. As a consequence of the choice of a vector-valued Reproducing Kernel Hilbert Space (RKHS) as a test space for integrating manifolds, it is possible to consider that shapes are embedded in this Hilbert Space. A vector-valued RKHS is a Hilbert space of vector fields; therefore, it is possible to compute a mean of shapes, or to calculate a distance between two manifolds. This embedding enables us to consider size-and-shape classification algorithms. \\
These algorithms are  applied to a
3D database obtained from an anthropometric survey of the
Spanish child population with a potential application to online sales of children's wear.

\textbf{keyword} Currents \and Statistical Shape Analysis \and Reproducing Kernel Hilbert Space  \and Children's body shapes  \and $k-$means.

\end{abstract}

\section{Introduction} \label{introduccion}

Many problems in medical imaging analysis and computer vision involve the
classification of objects based on their shapes or on their sizes and shapes. A significant amount of research and activity has been carried out in recent decades in the general area of shape analysis.

Throughout this work we are going to work with bodies, i.e.  geometrical objects with bounded boundaries. Several mathematical frameworks have been proposed in the literature to deal with such objects, three of these being  the most widely used.
Firstly, functions can be used to represent closed contours of  the objects (curves in 2D and surfaces in 3D);
secondly geometrical objects can also be treated as subsets of $\re^m$ and, finally,
these geometrical objects can be described as sequences of points
that are given by certain geometrical or anatomical properties (landmarks).
Shapes, in all these settings, are embedded into a space which is not a vector space (in a large number of cases it is a smooth manifold) and on which no natural metric is defined. This makes the definition of statistics particularly difficult; for example, there is no simple explicit way to compute a mean \citep{Pennec06}.

Recently we have considered the space of planar shapes, represented by simple closed plane curves  with a Sobolev-type metric \citep{Gualetal15}. This space has the property of being isometric to an infinite-dimensional Grassmann manifold of 2-dimensional subspaces, and we have used this isometry to compute geodesics, distances between shapes and mean shapes and we have applied these concepts in order to study different biomedical applications.
The corresponding theory for the shape space of surfaces was generalized by \citep{Baueretal11}.
However, these results consider parameterized curves and surfaces.

In this approach, the contour of each body (curve in $\re^2$, surface in $\re^3$, or hypersurface in $\re^n$), is represented by a mathematical structure named current, where unparameterized curves and surfaces are considered. This framework is not limited to a particular kind of data. Indeed, it provides a unifying framework to process any set of points, curves and surfaces or a mixture of these. No hypothesis on the topology of the shapes is assumed. In particular, it is robust to changes of connectivity of the structures. Moreover, it is weakly sensitive to the sampling of shapes and it  does not depend on the choice of  parameterization. However, the main advantage of this setting is that shapes are embedded into a vector space provided with an inner-product;  hence, it is possible to use easy statistical tools.

A current is a mathematical object which has been proved relevant for modeling geometrical data like curves and surfaces \citep{VaillantGlaunes05,Glaunesetal06,Durrlemanetal09}.

From integration on manifolds \citep{Morgan08,Lang95}, if $\Omega_{p}$ denotes the space of differential $p$-forms in $\mathbb{R}^{n}$, each p-dimensional submanifold $X$ in $\mathbb{R}^{n}$ (in particular, $X$ may be the contour of a geometrical object) can be represented by an application that integrates each p-form along $X$, i.e. by an application
\begin{eqnarray}\label{funcio1}
C_{X}:&&\Omega_{p}\longrightarrow \mathbb{R} \\
    &&w \longrightarrow \int_{X} w, \nonumber
\end{eqnarray}
such application $C_X$, is called a $p$-current.

In addition, it is possible to associate a subspace of currents to a Reproducing Kernel Hilbert Space (RKHS) by duality. A RKHS is a Hilbert space of mappings which has useful properties. Moreover, these associations allow us to represent each set of piecewise-defined manifolds by a function in a RKHS \citep{Durrleman10}.

Given a set of geometrical objects (curves or surfaces), our aim is to apply classification techniques developed for Euclidean spaces in order to divide the $m$ objects into $k$ appropriate clusters.
A Hilbert space and, even more so,  a RKHS can be considered the natural extension of the usual Euclidean spaces $\mathbb{R}^n$. The completeness of Hilbert spaces gives a framework in which to work with infinite-dimensional vectors as the limit of finite-dimensional vectors.

This paper arose as the result of an important study conducted by the Valencian Institute of Biomechanics,
 the ultimate objective of which was to help decision makers (parents/relatives/children) in the size selection process when shopping online for children's wear.

A 3D anthropometric study of the child population in Spain was carried ot for that purpose. After the study was completed, a database was generated consisting of $739$ randomly
selected Spanish children between $3$ and $12$ years of age.
They were scanned using the Vitus Smart 3D body scanner from Human
Solutions, a non-intrusive laser system
formed by four columns housing the optic system, which moves from head
to feet in ten seconds, performing a sweep of the body. Our work focuses on one of the aims of this study: to define of an efficient sizing system.

A standard sizing system classifies a specific population into
homogeneous subgroups based on certain key body dimensions
\citep{NormaUNE, ChungaetAl07, Ibanez2012}. Most of the
standard sizing charts propose sizes based on intervals over just
two or three anthropometric dimensions.
However, correlations between anthropometric measures show great variability in body proportion and as
a result it is not possible to cover so many different body morphologies with these kinds of models.

In this paper, instead of using clustering methods to divide the population into sizes by simply using a set of anthropometric variables, we propose to use the body shapes represented by currents and the well-known \emph{k-means} algorithm in the corresponding space.

\vspace{0.3cm}

The original \emph{k-means} algorithm \citep{Steinhaus56, Lloyd57}, endeavors to find a partition such that the
sum-of-squares error between the empirical mean of a cluster and the
objects in the cluster is minimized. It tries  to approximate
this optimum $k$-partition by iterating. Starting with $k$ arbitrary
initial cluster centers, an initial $k$-partition is obtained,
assigning each object to its closest cluster center. Next,
the new $k$ cluster centers are re-calculated as the mean of the observations of the
clusters resulting from the previous step. This loop is continued
until no further changes are made.
Many procedures were developed in subsequent decades
to improve this classic algorithm; see e.g. \cite{Kanungo2002} and \cite{Nazeer2009}.
Even though the \emph{k-means} algorithm was first proposed over $50$ years ago, it is still one of
the most widely used algorithms for clustering \citep{Jain2010}.

Our implementations have been written with \cite{Matlab14}.

The article is organized as follows:
Sections~\ref{teoria} and \ref{teoria2} concern the theoretical concepts  of currents and Reproducing Kernel Hilbert Spaces.
In section~\ref{k-means}  the \emph{k-means} algorithm in the RKHS space is introduced.
An experimental study with synthetic figures is conducted in Sections~\ref{exp_study2D} and ~\ref{exp_study3D}.
The application for classifying children's body shapes is detailed
in Section~\ref{our_appl}.
Finally, conclusions are discussed in Section~\ref{conclusions}.

\section{ From bodies to elements in a Reproducing Kernel Hilbert Space through currents } \label{teoria}

Let $B_1, B_2,\dots, B_m$ be  $m$ bodies in $\mathbb{R}^n$ whose boundaries $S_i = \partial B_i$, $(i=1,2,\dots,m)$, are smooth hypersurfaces in $\mathbb{R}^n$. In this section we introduce the theoretical foundations to represent $S_i$ as elements in a Reproducing Kernel Hilbert Space (RKHS). In order to do that, we will first represent the hypersurfaces $S_i$ as geometrical currents.

Currents were introduced by De Rham in 1955 and by the
1960  paper by  Federer and Fleming on `Normal
and Integral Currents'', which was awarded the 1986 AMS Steele Prize for a paper
of fundamental or lasting importance; but their use in computational anatomy is recent (J. Glaun\'es. PhD thesis. 2005).

Let $\Omega_{n-1}$ denote the space of continuous $(n-1)-$forms on $\re^n$. The space of $(n-1)-$currents on $\re^n$ is the topological dual $\Omega'_{n-1}$; i.e. the space of linear and continuous forms on $\Omega_{n-1}$; and it is a fact that every  hypersurface $S$ of $\re^n$ with a  finite volume can be represented by an element of $\Omega'_{n-1}$. That is, from integration on manifolds \citep{Morgan08}, we know that any $(n-1)-$dimensional form of $\Omega_{n-1}$ can be integrated along the hypersurface $S$, which associates with $S$ a $(n-1)-$current $C_S$ such that:
\begin{eqnarray}\label{funcio2}
C_S (\omega)=\int_S \omega,\qquad \forall \omega \in \Omega_{n-1}.
\end{eqnarray}
The application $S\longrightarrow C_S$ is injective but not surjective (not all the currents can be represented by integration on a hypersurface (geometrical currents)).

Suppose that the hypersurface $S$ is a parameterized surface $r: D\subset\re^{n-1} \longrightarrow \re^n$ with $r(D)=S$, then $r(x)=r(x_1,x_2, \dots,x_{n-1})\in S$, and
\begin{eqnarray}\label{funcio3}
C_S (\omega)=\int_S \omega = \int_D \omega (r(x)) \left( \frac{\partial r}{\partial x_1}\wedge\dots\wedge\frac{\partial r}{\partial x_{n-1}} \right)d x_1\dots d x_{n-1}.
\end{eqnarray}

This representation is fully geometric in the sense that it only depends on the hypersurface structure and not on the choice of parameterization. Moreover the representation of a surface as a geometrical current distinguishes between isometric hypersurfaces (that is, hypersurfaces obtained as rotations and/or translation). On the other hand, the opposite current $-C_S$  represents the same hypersurface but with the opposite orientation (since the flux through the hypersurface has the opposite sign in this case).\\

\subsection{Vectorial representation of geometrical currents}

A form $\omega\in \Omega_{n-1}$ can be associated with a vector field $\bar{\omega}$ on $\re^n$ thanks
to the isometric mapping between the $(n-1)$-form on $\Omega_{n-1}$ and the vectors on $\re^n$. Then,
\begin{eqnarray}\label{funcio4}
\begin{aligned} C_S (\omega)=& \int_D \omega (r(x)) \left( \frac{\partial r}{\partial x_1}\wedge\dots\wedge\frac{\partial r}{\partial x_{n-1}} \right)d x_1\dots d x_{n-1}\\=& \int_D   Det \left(\frac{\partial r}{\partial x_1},\dots,\frac{\partial r}{\partial x_{n-1}},\bar{\omega}(r(x)) \right)d x_1\dots d x_{n-1}.\end{aligned}
\end{eqnarray}

Formally, the association between forms and vectors is given by the Hodge star operator and duality \citep{doCarmo12}.

\subsection{Particular cases: planar curves and surfaces}

As stated in the introduction, we are  interested in the particular cases of planar closed curves and compact surfaces (contours of bodies in $\mathbb{R}^2$ or $\mathbb{R}^3$).\\

Let $\alpha: I=[a,b]\longrightarrow \re^2$ be a parameterized regular oriented simple curve in $\re^2$. We associate with $\alpha$ the function (geometrical current)
\begin{eqnarray}\label{funcio5}
C_{\alpha}(\omega)=\int_a^b \bar{\omega} (\alpha(t))\cdot \alpha' (t)\, dt,
\end{eqnarray}
where $\bar{\omega}$ is a vector field in $\re^2$ and $\cdot$ denotes the inner product in $\mathbb{R}^2$.

Let $S$ be an orientable parameterized surface in $\re^3$ given by  $r:U\subset\re^2\longrightarrow \re^3$; that is, $r(U)=S$. We associate with $S$ the function (geometrical current)

\begin{eqnarray}\label{funcio6}
C_{S}(\omega)=\int_U \bar{\omega} (r(u,v))\cdot (r_u (u,v)\wedge r_v (u,v))\, du\, dv,
\end{eqnarray}
where $\bar{\omega}$ is a vector field in $\re^3$, $\cdot$ denotes the inner product in $\mathbb{R}^3$, $r_u =\partial r / \partial u$ and $r_v =\partial r / \partial v$.\\

Then, to characterize hypersurfaces (mainly curves and surfaces) from the above expressions, we measure how these integrals vary as the vector field $\omega$ varies. However, instead of considering all the vector fields, we will define a test space of square-integrable vector fields where $\omega$ varies. In particular, as in \cite{Durrleman10}, we will choose as the test space  a vector-valued Reproducing Kernel Hilbert Space (RKHS).

It is important to note that in this case the application $S\longrightarrow C_S$ will not be  injective; that is, the same geometrical current $C_S$, as a map defined in a  RKHS, may represent two different hypersurfaces. Therefore, the choice of the appropriate RKHS will depend on the application proposed.

\subsection{Operator-valued kernels and test space vector-valued RKHS.}

This section gives a definition of  a vector-valued Reproducing Kernel Hilbert Space  from the Riesz representation theorem  and studies its properties.

The abstract theory of scalar-valued RKHS was developed by \cite{Aronszajn50}. A scalar-valued RKHS is a  Hilbert space of functions $f:\mathbb{R}^n \rightarrow  \mathbb{R}$ with some practical properties. In recent years years, the study of RKHS has been extended to vector-valued functions (the space contains vector fields from $\mathbb{R}^n$ to $\mathbb{R}^n$) (see \cite{Carmelietal06}, \cite{Micchellietal05} and \cite{Caponnettoetal08}) and it has now become a widely studied theory . \\

Let $\mathcal{L}(\mathbb{R}^{n})$ be the Banach space of bounded linear functions from $\mathbb{R}^{n}$ to $\mathbb{R}^{n}$.

\begin{defn} Let $W$ be a Hilbert space of vector fields from $\mathbb{R}^{n}$ to $\mathbb{R}^{n}$. An operator $K:\mathbb{R}^n \times \mathbb{R}^n \rightarrow \mathcal{L} (\mathbb{R}^n)$ is said to be an operator-valued reproducing kernel (rk) associated with $W$ if
\begin{enumerate}
\item for every $x, \alpha \in \mathbb{R}^n$, $K(\cdot, x) (\alpha) \in W$  (where  $K(\cdot, x) (\alpha)(y)= K(x,y)(\alpha) \in \mathbb{R}^n$, $\forall y \in \mathbb{R}^n$) and,

\item $K$ satisfies the "reproducing property"; that is, $ \forall \omega \in W$ and $x, \alpha \in \mathbb{R}^n$
\begin{eqnarray}\label{funcio7}
\omega(x) \cdot \alpha= \langle K( \cdot , x) (\alpha), \omega \rangle_W  \nonumber
\end{eqnarray}
\end{enumerate}
\end{defn}

\begin{defn} Let $W$ be a Hilbert space of vector fields from $\mathbb{R}^{n}$ to $\mathbb{R}^{n}$. $W$ is a vector-valued RKHS if there is an operator-valued reproducing kernel (rk) associated with $W$.
\end{defn}


 The next theorem (see below) is a sort of converse to this: if a function $K$ is both symmetric and positive definite,  then there is a Hilbert space of vector fields from $\mathbb{R}^{n}$ to $\mathbb{R}^{n}$ for which it is a reproducing kernel.

\begin{defn} A function $K:\re^n\times\re^n\longrightarrow{\cal L}(\re^n)$ is said to be an operator-valued positive definite and self-adjoint kernel if for each pair $(x,y)\in\re^n\times\re^n$, $K(x,y)\in {\cal L}(\re^n)$ is a self-adjoint operator and
$$\sum_{i,j=1}^N \alpha_i \cdot K(x_i,x_j)(\alpha_j)\ge 0,$$
for every finite set of points $\{x_i\}_{i=1}^N$ in $\re^n$ and $\{\alpha_i\}_{i=1}^N$ in $\re^n$.
\end{defn}

\begin{thm}
 If K is an operator-valued positive definite and self-adjoint kernel, then there is a unique RKHS, $W$, such that $K$ is the operator-valued reproducing kernel (rk) associated with $W$.
\end{thm}

It is essential to bear in mind that the proof of this theorem is based on constructing the space $W$ through the completion of $H_0:=span\{K(\cdot, x)(\alpha)\,\,  / \,\, x,  \alpha  \in \mathbb{R}^{n}\}\subset W$.

For $\omega_{1}=\displaystyle \sum_{i=1}^{N_1}K(\cdot,x_{i})(\alpha_{i})$, $\omega_{2}=\displaystyle \sum_{j=1}^{N_2}K( \cdot, x_{j})(\beta_{j}) \in H_{0}$, define
\begin{eqnarray}\label{funcio8}
\langle \omega_{1}, \omega_{2} \rangle_{H_{0}}:= \displaystyle \sum_{i=1}^{N_1} \displaystyle \sum_{j=1}^{N_2} \alpha_{i} \cdot K(x_{i}, y_{j})(\beta_{j}).
\end{eqnarray}
The vector-valued RKHS $W$ associated with the kernel $K$, that, from now on, will be denoted as $H_K$, is the closure of $H_{0}$, that is, the span of vector fields of the form
$K(x, \cdot)(\alpha)$ for every $x\in\re^n$ and $\alpha\in\re^n$ is dense in $H_K$. For this reason, the inner product in $H_K$ between $\omega_1 , \omega_2 \in \overline{H_0}$ is the limit of expression (\ref{funcio8}) when $N_1$, $N_2$ tend to infinity.\\

Having established the vector-valued RKHS test space $H_K$, currents will be evaluated in $H_K$; that is,  the space of  currents  considered is $H_{K}^{*}$  (it contains the continuous linear functions from $H_{K}$ to $\mathbb{R}$) which includes the geometrical currents as a  subset. The space of  currents  $H_{K}^{*}$ is a vector space with the operations sum ($+$) and product ($\cdot$) as in a standard space of functions \citep{Durrleman10}.  \\

Therefore, the idea is to build the vector space spanned by the vector fields of the form $K(x, .)(\alpha)$
and to make this space complete by adding the limit of every Cauchy sequence to it.
This construction make it possible  to process  discrete meshes of surfaces
 and continuous surfaces (limit of such a finite
combination) in the same setting.

\subsection{Curves and Surfaces as elements in a vector-valued RKHS}

We are now going to use the properties of the RKHS, $H_K$ in order to rewrite the geometrical currents associated with curves in $\re^2$ and surfaces in $\re^3$.\\

Let $\omega$ be a vector field in $H_K(\mathbb{R}^2, \mathbb{R}^2)$. Then, by using  Eq. (\ref{funcio5}), the geometrical current associated with a curve $\alpha$ becomes:
\begin{eqnarray}\label{funcio9}
C_{\alpha}^{K}(\omega)=\int_a^b \omega(\alpha(t))\cdot \alpha'(t)\, dt = \langle \int_a^b K(\alpha(t), \cdot) (\alpha'(t)) dt, \omega \rangle_{H_K},
\end{eqnarray}
and by Eq. (\ref{funcio6}), the geometrical current associated with the parameterized surface $S=r(U)$ is:
\begin{eqnarray}\label{funcio10}
 C_{S}^{K}(\omega)=\int_U \bar{\omega} (r(x))\cdot \eta(x)\, dx= \langle\int_U K(r(x), \cdot)  (\eta(x))\, dx, \omega \rangle_{H_K}
\end{eqnarray}
where $x=(u,v)$, $dx=du\, dv$,  $\eta(x)=r_{u}\wedge r_{v}$ is the orthogonal vector  to the surface $S$ at the point $r(x)$, and $\omega$ and $\int_{S} K(r(x), \cdot)(\eta(x)) dx \in H_K(\mathbb{R}^3, \mathbb{R}^3)$.

Until now, each hypersurface has been associated with an element in the vector space $H_{K}^{*}$.
However, the Riesz Fr\`echet Theorem (\cite{Conway13}) establishes that there is an isometric, linear, bijective mapping  $\mathcal{L}_{H_K}:H_K\rightarrow H_{K}^{*}$, defined by $\mathcal{L}_{H_K}(\omega)(\omega'):=\langle \omega, \omega' \rangle_{H_K}$, $\forall \omega, \omega' \in H_K$. As a consequence, the space $H_{K}^{*}$ is isometric  to $H_{K}$, and  then  each hypersurface can be associated with  a vector field in $H_K$.

Therefore, as a result of the Riesz-Fr\`echet Theorem, it is possible to represent the parameterized curve $\alpha$ defined in $[a,b]$, by an element in $H_K(\mathbb{R}^2, \mathbb{R}^2)$; that is

\begin{eqnarray}\label{funcio11}
\alpha\longrightarrow C_{\alpha}^{K}(\omega)= \langle \int_a^b K(\alpha(t), \cdot) (\alpha'(t)) dt, \omega \rangle_{H_K}\cong \int_a^b K(\alpha(t), \cdot) (\alpha'(t)) dt,
\end{eqnarray}
where $\cong$ denotes the isometric element from the Riesz-Fr\`echet Theorem, and to represent a parameterized surface $S$ by
\begin{eqnarray}\label{funcio14}
S \longrightarrow C_{S}^{K}(\omega)=\int_U \bar{\omega} (r(x))\cdot \eta(x)\, dx= \langle\int_U K(r(x), \cdot)  (\eta(x))\, dx, \omega \rangle_{H_K}\cong \int_U K(r(x), \cdot)  (\eta(x))\, dx
\end{eqnarray}

Consider now that the curve $\alpha$ is only known at a finite number, $p$, of points $\{t_1 < t_2 < \ldots, <t_p\}$, that constitute a partition of the interval $[a,b]$. Let  $y_j = \alpha(t_j) \in \mathbb{R}^2$ $\forall j=1,\cdots ,p$ , let  $x_j\in\re^2$ denote the center of the segment $[y_j, y_{j+1}]$ and let $\tau_j=y_{j+1}-y_j$ be an approximation of the tangent vector (the finer the partition the better). Then,
\begin{eqnarray}\label{funcio12}
\alpha\longrightarrow C_{\alpha}^{K}(\omega) \cong \int_a^b K(\alpha(t), \cdot) (\alpha'(t)) dt = \displaystyle \lim_{p\rightarrow \infty} \displaystyle \sum_{j=1}^{p}K(x_j, \cdot)(\tau_j) \sim \displaystyle \sum_{j=1}^{p}K(x_j, \cdot)(\tau_j).
\end{eqnarray}

In practical applications each curve $\alpha$ will be represented by a finite addition which  is an approximation to the vector field.

\begin{figure}
\begin{center}
\begin{tabular}{c}
\includegraphics[width=10cm]{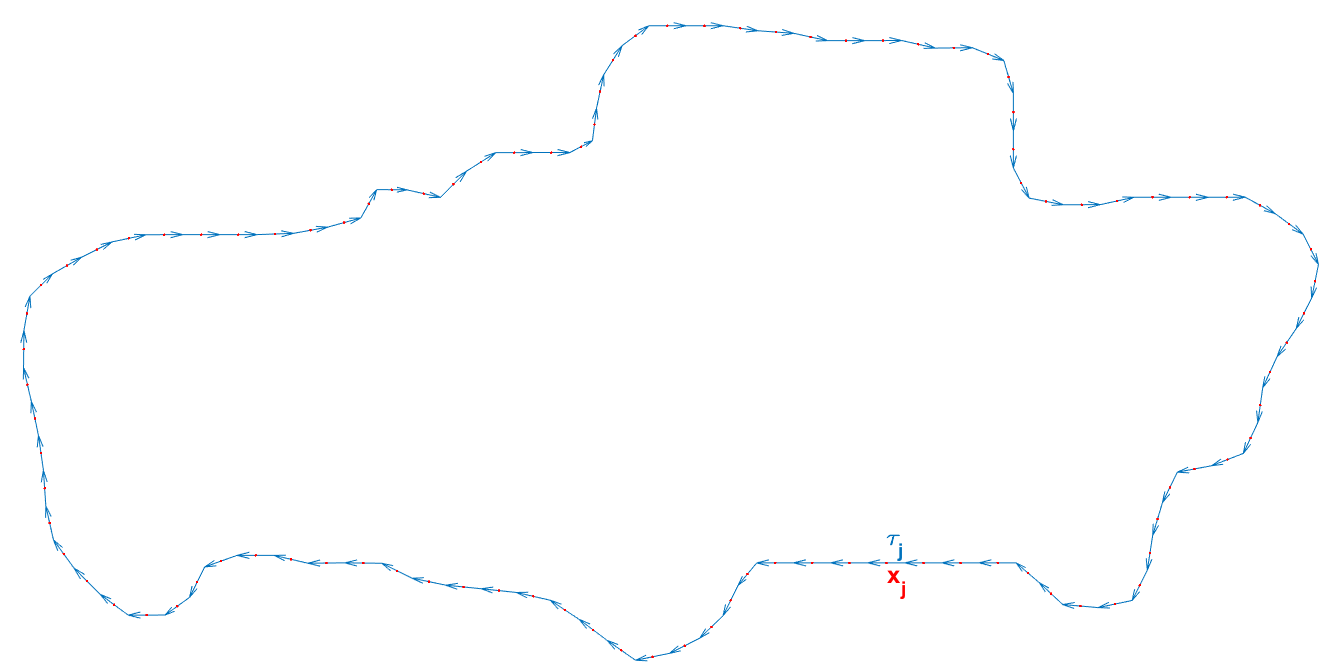}
\end{tabular}
\caption{Curve in $\mathbb{R}^2$ and its elements}
\end{center}
\end{figure}

 If we suppose we have a triangulation of $S$ where each triangle $y_{j}y_{j+1}y_{j+2}$ is represented by the vector field $K(x_{j}, \cdot)(\tau_{j})$
 where $x_{j}=\frac{1}{3}(y_{j}+y_{j+1}+y_{j+2})$ and $\tau_{j}=\frac{1}{2}(y_{j+1}-y_{j})\wedge (y_{j+2}-y{j})$  ($\tau_{j}$ is the normal vector to the triangle,
 whose norm encodes the area of the triangle); then,
\begin{eqnarray}\label{funcio15}
S \longrightarrow C_{S}^{K}(\omega) \cong \int_U K(r(x), \cdot)  (\eta(x))\, dx \sim \displaystyle \sum_{j=1}^{p}K(x_j, \cdot)(\tau_j).
\end{eqnarray}

 The finite addition tends towards the integral as the mesh is refined. This finite addition is an approximation to the vector field $\int_{S} K(r(x), \cdot)(\eta(x)) dx$, and it will be the representation which will be used as a consequence of its computational simplicity.

\begin{figure}
\begin{center}
\begin{tabular}{c}
\includegraphics[width=10cm]{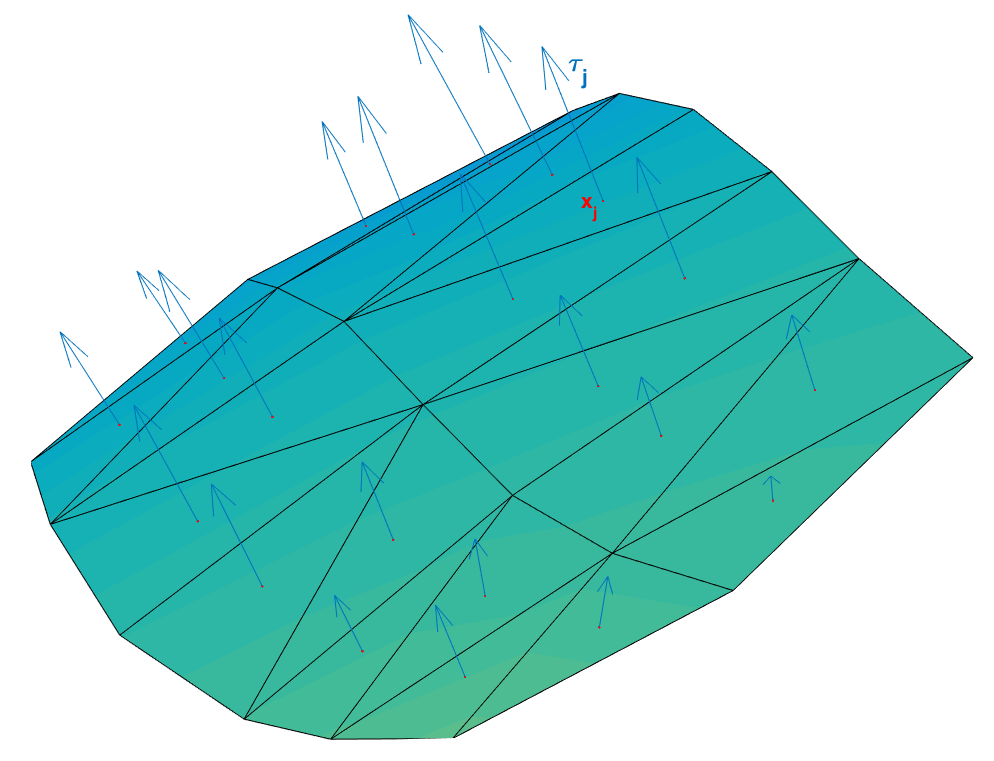}
\end{tabular}
\caption{Triangulated surface in $\mathbb{R}^3$ and its elements}
\end{center}
\end{figure}

Consequently, we deal with curves and surfaces as  vector fields in $H_K$. Thus, the distance between two surfaces (or curves) is defined as the distance between the corresponding elements in $H_k$; that is,  if $\varphi_1$ and $\varphi_2$ are two  elements in the RKHS associated with two surfaces $S_1$ and $S_2$; then,
\begin{eqnarray}\label{funcio16}
d(S_1, S_2)^2= d(\varphi_1, \varphi_2)^2=\langle \varphi_1- \varphi_2, \varphi_1-\varphi_2 \rangle_{H_{K}}=\langle \varphi_1, \varphi_1\rangle_{H_{K}}-2\langle \varphi_1, \varphi_2\rangle_{H_{K}}+\langle \varphi_2, \varphi_2\rangle_{H_{K}}
\end{eqnarray}
Using finite approximations to   $\varphi_1=\displaystyle\sum_{j=1}^{p} K(x^1_{j}, \cdot)(\tau^1_{j})$ and   $\varphi_2=\displaystyle\sum_{j=1}^{p} K(x^2_{j}, \cdot)(\tau^2_{j})$ we have
\begin{eqnarray}\label{funcio17}
 \langle \varphi_1, \varphi_2 \rangle _{H_{K}}= \displaystyle \sum_{i=1}^{p} \displaystyle \sum_{j=1}^{p}  \tau^1_{i} \cdot K(x_{i}^1, x_{j}^2)(\tau^2_{j}).
 \end{eqnarray}

Moreover, the sample mean in the vector-valued RKHS $H_K$ is calculated in the same way as in Euclidean spaces \citep{HsingEubank15}. Given a sample $\varphi_1,..., \varphi_m$ the sample mean is:
\begin{eqnarray}\label{funcio18}
 \bar{\varphi}=\frac{ \sum_{l=1}^m \varphi_l}{m}=  \frac{\left(  \displaystyle \sum_{l=1}^m \sum_{j=1}^{p} K(x^l_{j}, \cdot)(\tau^l_{j}) \right)}{m},
  \end{eqnarray}
$ \bar{\varphi}\in H_K$; however, in general, $ \bar{\varphi}$ is not a geometrical current associated with a surface. \\

 The distance between surfaces, obtained from the distance between the corresponding geometrical currents,  gives  a  global estimate of the shape dissimilarity between objects. This distance will be used for the definition of an efficient sizing system for the child population, in which a global dissimilarity shape measure between bodies that does not highlight where  the differences occur locally is needed.

\section{Choice of the operator-valued reproducing kernel}\label{teoria2}
The choice of the kernel determines the vector-valued RKHS
and especially its metric. The choice of this metric is therefore crucial and must be
adapted to every particular application.

Based on this, we are going to use a particular class of operator-valued kernel $K$ defined as follow:

\begin{defn}\label{def1}
Let $D \subseteq \mathbb{R}^{n}$ be a non-empty subset, and $K: D \times D \rightarrow \mathcal{L}(\mathbb{R}^n)$ an application.
Then, for each $x, y \in D$, the operator $K(x,y)$ is defined by
\begin{eqnarray}\label{funcio20}
K(x,y):&&\mathbb{R}^{n} \longrightarrow \mathbb{R}^{n} \\
    && \alpha \longrightarrow K(x,y)(\alpha):=k(x,y)\cdot  \alpha \nonumber
\end{eqnarray}
where $k: D \times D \rightarrow \mathbb{R}$ is a symmetric and positive semi-defite function, (i.e. $k(x,y)=k(y,x)$ $\forall x, y \in  D$ and $\displaystyle \sum_{i,j} a_i k(x_i, x_j)(a_j)\geq 0$ for finite sets $\{a_i\}\in \mathbb{R}$, $\{x_i\}\in D$).
\end{defn}

\begin{prop}
The operator-valued $K$ established in the previous definition is well defined, symmetric and positive semi-definite, so there is a unique RKHS vector-valued $H_K(D, \mathbb{R}^n) \subset W:= \{ \omega: D \longrightarrow \mathbb{R}^n \}$ (or simply $H_K$) with $K$  as its rk.
\end{prop}

{\it Proof.}
Given $(x,y) \in D \times D$, the operator $K(x,y)$ is obviously linear. In addition, $K(x,y)$ is bounded because its norm is bounded by  $|k(x,y)|$:

\begin{eqnarray}\label{funcio21}
\| K(x,y) \|=\sup \{ \|K(x,y)(\alpha) \|_{\mathbb{R}^{n}} : \alpha \in \mathbb{R}^{n},  \|\alpha \|_{\mathbb{R}^{n}}\leq 1 \} \leq | k(x,y) |.
\end{eqnarray}

Moreover, as $k$ is symmetric and positive semi-definite,  $K$ immediately has these properties.
Hence, $K$ is an operator-valued reproducing kernel and there is a unique RKHS vector-valued $H_K(D, \mathbb{R}^n)$ with $K$ as its rk. $\square$\\

Although it is not known how to choose the ``best'' kernel for a given application, translation-invariant isotropic scalar
kernels of the form $k(x,y)=k(\| x- y \|_{\mathbb{R}^n})$ are often used. In particular the Gaussian function (also called the Gaussian kernel)
\begin{eqnarray}\label{funcio22}
 k(x,y):= \displaystyle e^{\displaystyle \frac{-\| x- y \|^2_{\mathbb{R}^n} }{\lambda^2}} .
\end{eqnarray}
where $\lambda > 0$ is a scale parameter  (bandwidth), defines an operator-valued $K:D \times D \rightarrow \mathcal{L}(\mathbb{R}^n)$ that is  particularly important in the literature, called Vector-Valued Gaussian, Kernel which fixes a vector-valued RKHS $H_K(D, \mathbb{R}^n)$ with $K$ as its rk.
This fixed vector-valued RKHS  has the following expression (\cite{Quangetal10})
\begin{eqnarray}\label{funcio23}
H_K(D, \mathbb{R}^n)=& & \lbrace f \in C_{0}(D, \mathbb{R}^{n})\cap L^{2}(D, \mathbb{R}^{n}) : \int_{\mathbb{R}^n} e^{\frac{\lambda^{2}\|\xi\|_{\mathbb{R}^n}^{2}}{4}}\| \widehat{f}(\xi)\|_{\mathbb{R}^n}^{2} d\xi < \infty \rbrace
\end{eqnarray}
where $\widehat{f}$ is the Fourier transform of $f$.

\begin{prop}\label{prop1}
Given $D \subseteq \mathbb{R}^{n}$  a non-empty subset;  $K_{\lambda}: D \times D \rightarrow \mathcal{L}(\mathbb{R}^n)$ the operator defined as in Eq.(\ref{funcio20}) through the Gaussian Kernel $k$ with parameter $\lambda$ (Eq. \ref{funcio22}), and $H_{K_{\lambda}}$ the RKHS established by $K_{\lambda}$, if $ \lambda_1 , \lambda_2 \in \mathbb{R}:$ $0< \lambda_1 \leq \lambda_2$, then  $H_{K_{\lambda_2}} \subseteq H_{K_{\lambda_1}}$.
\end{prop}

{\it Proof.}
If $f$ was in $H_{K_{\lambda_2}}$, using the space expression (\ref{funcio23}), the integral with $\lambda_2$ would be finite and larger than the integral with parameter $\lambda_1$,

\begin{eqnarray}\label{funcio24} \nonumber
\int_{\mathbb{R}^n} e^{\frac{\lambda_1^{2}\|\xi\|^{2}}{4}}\| \widehat{f}(\xi)\|^{2} d\xi \leq \int_{\mathbb{R}^n} e^{\frac{\lambda_2^{2}\|\xi\|^{2}}{4}}\| \widehat{f}(\xi)\|^{2} d\xi < \infty  \nonumber
\end{eqnarray}

Hence, $f \in H_{K_{\lambda_1}}$. In conclusion, the smaller the value of $ \lambda > $ 0 , the greater the space $H_{K _ { \lambda } }$ established. $\square$

\begin{rem} \textbf{Choice of parameter $\lambda$}\end{rem}
Since, the smaller the value of $ \lambda > $ 0 , the greater the space $H_{K _ { \lambda } } $ established, (proposition \ref{prop1}). It is important to note that the larger $H_{K_{\lambda}}$, the better the differentiation between geometrical data, because since the test space would be greater, it would be more likely to identify different values that the current has in vector fields. Thus, the smaller the value of $ \lambda > 0$, the greater the  precision for  characterizing geometrical data.
For this reason, if $\lambda$ is too small, the distance in RKHS detects tiny geometrical details and too much noise could  be captured.
In conclusion, it is essential to choose a suitable parameter balancing the two previous ideas, so $\lambda > 0$ should be the typical scale at which the vector fields $\omega \in H_K(D, \mathbb{R}^{n})$ may vary spatially.

\section{k-means algorithm in the RKHS space} \label{k-means}
In this section we review the classic \emph{k-means} partitioning algorithm and we comment how adapt it to the RKHS space introduced in the previous section.

Given $m$ data points $x_1, ..., x_m$
in $\re^p$  and a $k$-partition $\mathcal{C}= (C_1, \ldots , C_k )$ of the set $\mathcal{O} = \{1, \ldots, m\}$ of underlying
objects, with non-empty classes, let:

\begin{equation} \label{objetivo1}
 W(\mathcal{C})=\sum_{i=1}^k\sum_{l \in C_i}\|x_l-\bar{x}_i\|^2,
\end{equation}
where $\bar{x}_i$ denotes the centroid of the data points.

As is well known, the classic k-means clustering approach looks for a $k$-partition $\mathcal{C}$ of $\mathcal{O}$ such that a minimum value of $W(\mathcal{C})$ is reached. This one-parameter optimization problem is equivalent to the two-parameter optimization problem:

\begin{equation} \label{objetivo2}
W(\mathcal{C}, z_1, \ldots,z_k)=\sum_{i=1}^k\sum_{l \in C_i}\|x_l-z_i\|^2,
\end{equation}
where minimization is also w.r.t. all vectors $Z = (z_1, \ldots, z_k )$ of $k$ points
$z_1, \ldots, z_k$ from $\re^p$ (class representatives, class prototypes).

The \emph{k-means} algorithm tries to approximate an optimum $k$-partition by iterating
the partial minimization steps \citep{Bock2007}:
\begin{algorithm} \nonumber
\begin{itemize}
\item \textbf{STEP 1.} Given an initial partition $\mathcal{C}(0)$, obtain the centroid vector $Z(0)=(\bar{x}_1^0, \ldots, \bar{x}_k^0)$. Set $i=1$.
\item\textbf{STEP 2.} Given a centroid vector $Z(i-1)$, obtain $\mathcal{C}(i)$ minimizing Eq.(\ref{objetivo2}) with respect to $\mathcal{C}$, assigning each point to the class whose centroid has the minimum Euclidean distance to it;
\item\textbf{STEP 3.} Given $\mathcal{C}(i)$, minimize  Eq.(\ref{objetivo2}) with respect to $Z$, obtaining the new centroid vector $Z(i)=(\bar{x}_1^{i}, \ldots, \bar{x}_k^{i})$, the sample means.
\item\textbf{STEP 4.} Set $i=i+1$ and go to STEP 2 until convergence is reached.
\end{itemize}
\end{algorithm}
By construction, this algorithm yields a sequence $Z(0), \mathcal{C}(1), Z(1), \mathcal{C}(2), Z(2), ... $ of centroids and partitions with decreasing values of the objective function (Eq. \ref{objetivo2}) that typically converges towards a (typically local) minimum value.

The new centroid vector obtained at each STEP 3 of this algorithm decreases the value of the objective function because of the fact that  the sample mean minimizes the Euclidean distance of any point in the cluster.

In this paper, our sample is a set of vector fields $\varphi_j= \displaystyle \sum_{i=1}^ p K(x_i^j, \cdot)(\tau_i^j) \in H_K$ ($j=1, \ldots, m$) which represent geometrical data by currents.
The adaptation of the above algorithm to the RKHS space of these vector fields is straightforward.

In this space the proximity between elements in the sample (Euclidean distance in equations \ref{objetivo1} and \ref{objetivo2}) is measured through distance of vector fields in $H_K$.
Thus, in the functions to minimize in the previous Algorithm, the Euclidean distance must be replaced by the distance given in equation \ref{funcio16}.

The sample means of STEP 3 in the previous algorithm are obtained using Eq. \ref{funcio18}.

\section{Experimental 2D Study} \label{exp_study2D}
In this section we study the performance of our procedure
in a shape classification problem using a known
database of synthetic figures called the
MPEG7CEShape-1 PartB database. It includes binary images grouped into categories like cars, faces, watches, horses and  birds with images corresponding to the same item, but showing noticeably different shapes.

To perform this experimental study, three classes from this database of
synthetic figures were considered: cars, faces and watches. Each class contains 20 elements (binary images) except the watch class where two of them were rejected (watch-2 is an atypical element because of its very large  size and watch-8 is turned and our theoretical framework considers size and shape). The figures were centered and the contour $\alpha_k$ from each of them defined an oriented smooth curve which was discretized by $100$ points $\{ y_j^k\}_{j=1}^{100}$  ($y_1^k=y_{100}^k$) for $k=1, \ldots, 58$. Moreover, face figures were rotated by $90$ degrees, in order to keep a common horizontal orientation in all synthetic figures (establishing the correspondence with children database, where all the elements are registered and have the same position). For each $k\in \{1, \ldots, 58\}$, from $\{ y_j^k\}_{j=1}^{100}$, we defined $x_j^k=\displaystyle \frac{1}{2} (y_j^k + y_{j+1}^k)$ the centers of the segments and the vectors $\tau_j^k=y_{j+1}^k-y_j^k$, $\forall j=1, \ldots, 99$,  which define the vector field $\displaystyle \sum_{j=1}^{99} K(x_j^k, \cdot)(\tau_j^k)$ in $H_K(D, \mathbb{R}^2)$.

\begin{figure}
\begin{center}
\begin{tabular}{c}
\includegraphics[width=10cm]{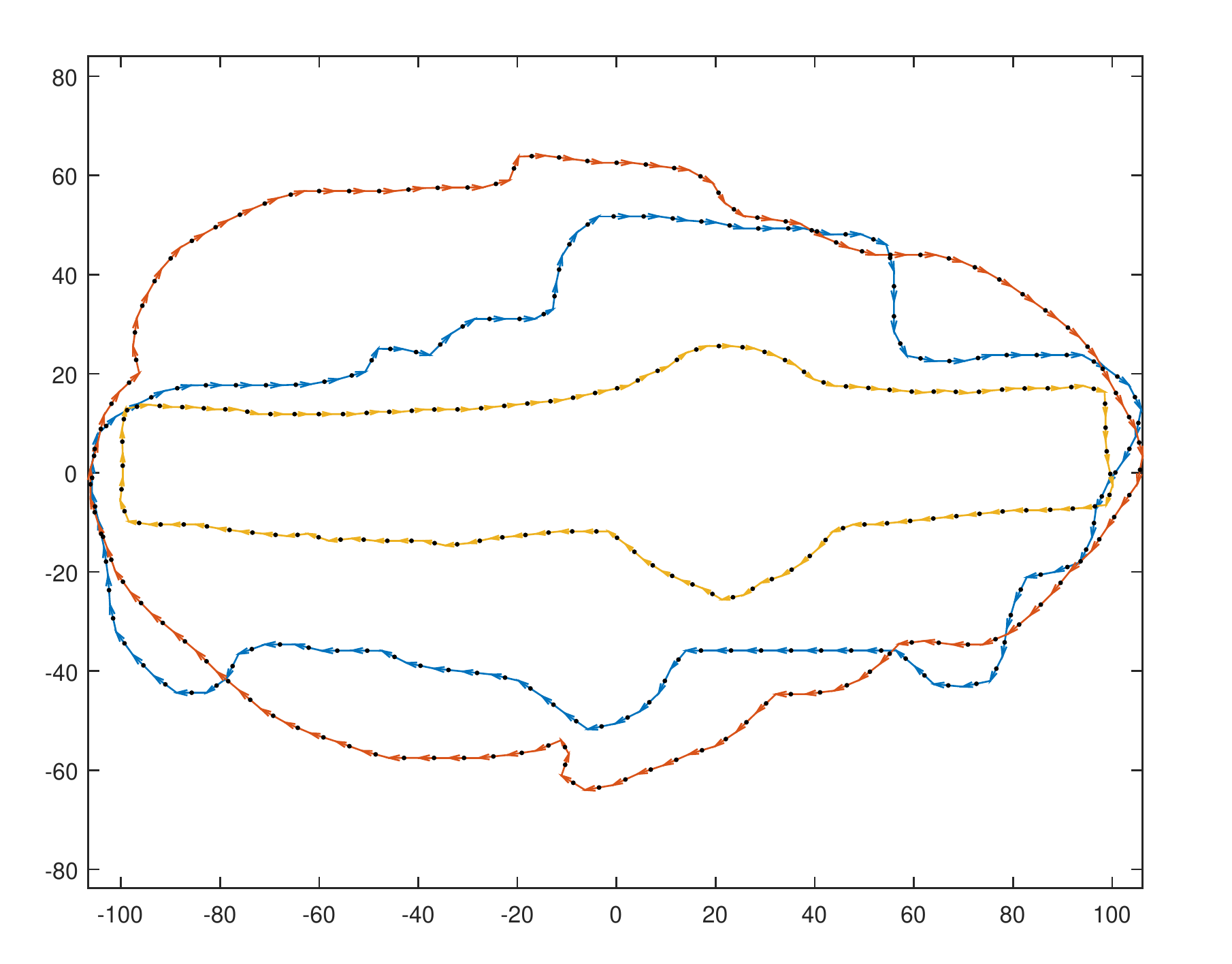}
\end{tabular}
\caption{An object from each class from synthetic 2D database (scenario 1). \label{figuras_ejemplo1}}
\end{center}
\end{figure}

In this experimental study, we were interested in analyzing similar situations to the ones that will appear in our real application in Section \ref{our_appl}. We therefore considered two scenarios. In the first, all the $58$ synthetic figures were contracted or expanded to reach the same length in the $X$ axis, establishing the similarity between this length and the height of a child.
Fig. (\ref{figuras_ejemplo1}) shows an example of an object from each class in this first situation. The points $\{ x_j^k \}_{j=1}^{99}$ are plotted in black and the vectors  $\{ \tau_j^k \}_{j=1}^{99}$  from each curve are plotted in different colors.

In the second scenario, half of the synthetic figures from each category were enlarged by a scale factor of  1.5. In this case there were two different "heights" for each class of figures. Moreover, each figure of the sample was multiplied by a random coefficient ranging between 1 and 1.1, in order to change the "height" of the figures somewhat.

\begin{figure}
\begin{center}
\begin{tabular}{c}
\includegraphics[width=10cm]{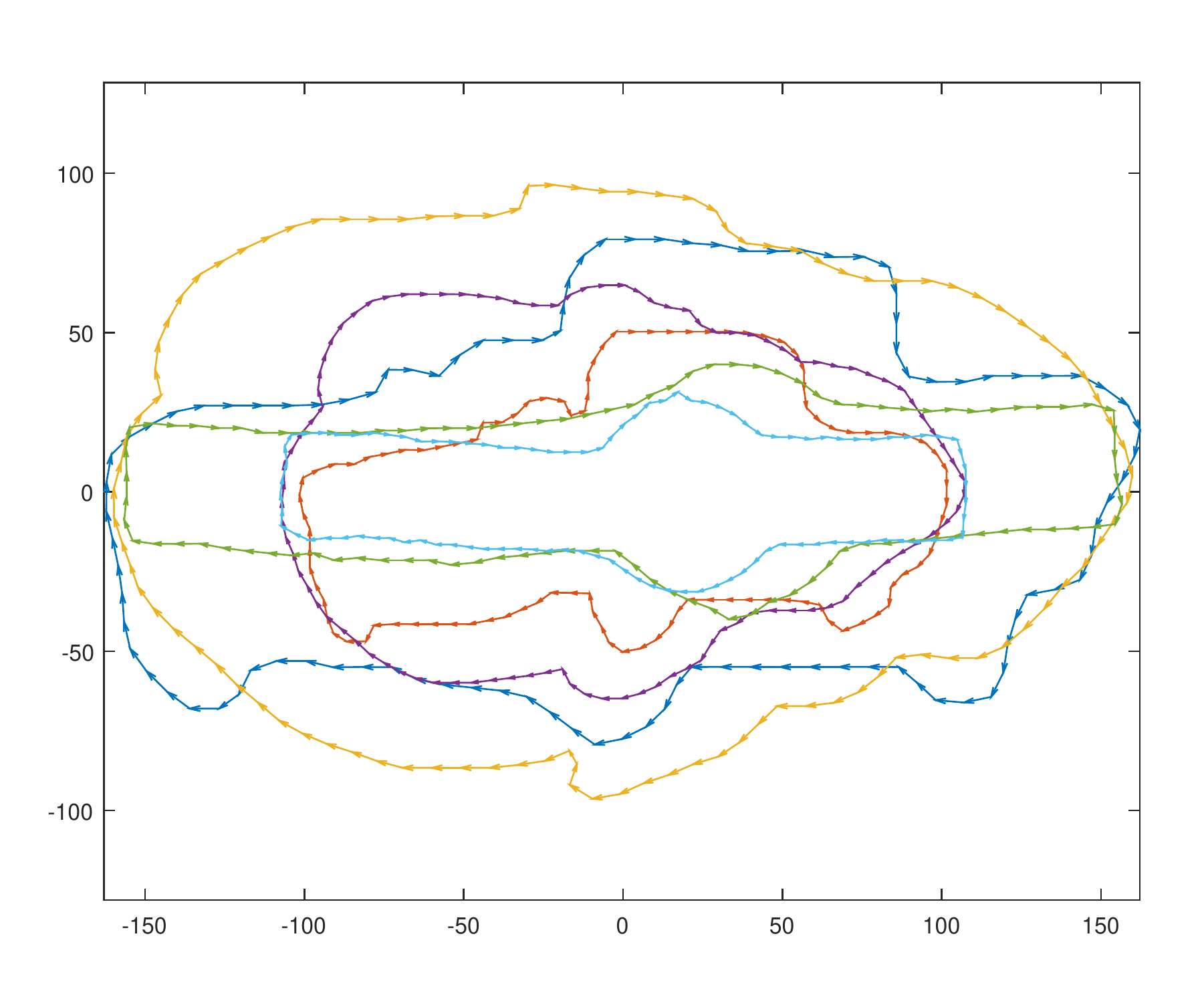}
\end{tabular}
\caption{An object from each class from the 2D synthetic database (scenario 2). \label{figuras_ejemplo2}}
\end{center}
\end{figure}

Fig. (\ref{figuras_ejemplo2}) shows an example
of an object from each group in the second situation, in which there are two "heights" from each class of shapes.

The \emph{k-means} algorithm presented in section \ref{k-means} was then applied to both scenarios,  choosing as the value of the parameter $\lambda$ the standard deviation of the points $\{ x_j^k\}_{j=1}^{99}$, $k=1, \ldots, 58$ which define the curves of the sample of each scenario. So we defined the Gaussian kernels with $\lambda=50.16$ in the first scenario and with $\lambda=67.35$ in the second.
In the first scenario the k-means algorithm recovered the three groups in accordance with the categories of figures in the database.
In the second scenario,  the $k$-means algorithm with $k$=6, grouped together the figures of the database with the same "height" and shape, as hoped.

\section{Experimental 3D Study} \label{exp_study3D}
In this section we describe the experiments carried out with a database of 3D figures, to test the same situations as in the previous 2D database. Now three classes of 3D objects: ellipsoids, spheres and pears are considered. Each class contains 10 elements centered in the origin and the contour from each of them is defined by an oriented smooth triangulated surface immersed in $\mathbb{R}^3$. Each surface $k$, is defined from $10000$ triangles with barycenter $x_{j}^k$ and normal vectors $\tau_{j}^k$, $j=1,\ldots, 10000$, being $\displaystyle \sum_{j=1}^{10000} K(x_j^k, \cdot)(\tau_j^k)$ in $H_K(D, \mathbb{R}^3)$ the vector field associated with each surface.

\begin{figure}
\begin{center}
\begin{tabular}{c}
\includegraphics[width=7cm]{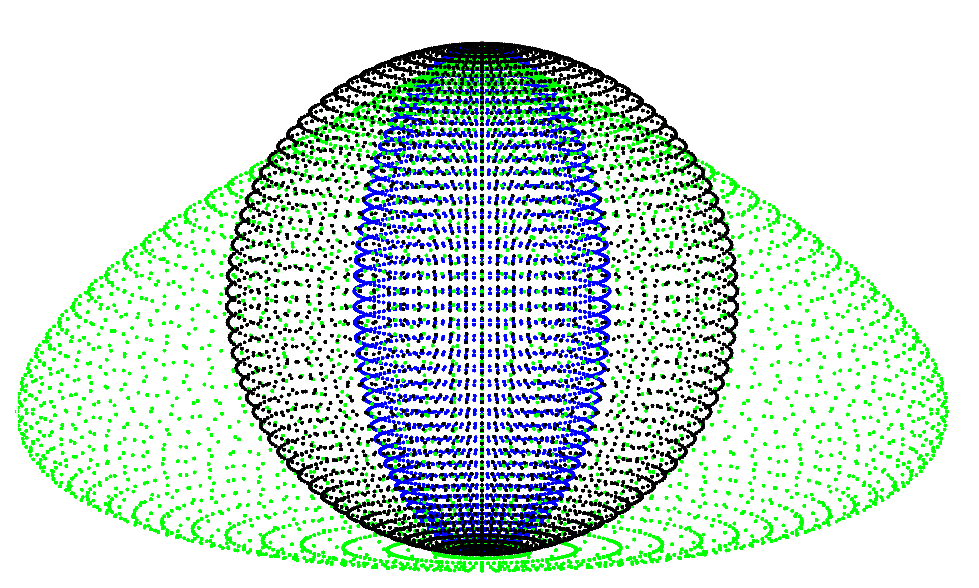}
\end{tabular}
\caption{An object from each class from the 3D synthetic database (scenario 1). \label{figuras_ejemplo11}}
\end{center}
\end{figure}

We consider the same scenarios as above. In the first, (Fig. (\ref{figuras_ejemplo11})) all figures have approximately the same length in the Y axis (same "height").

\begin{figure}
\begin{center}
\begin{tabular}{c}
\includegraphics[width=7cm]{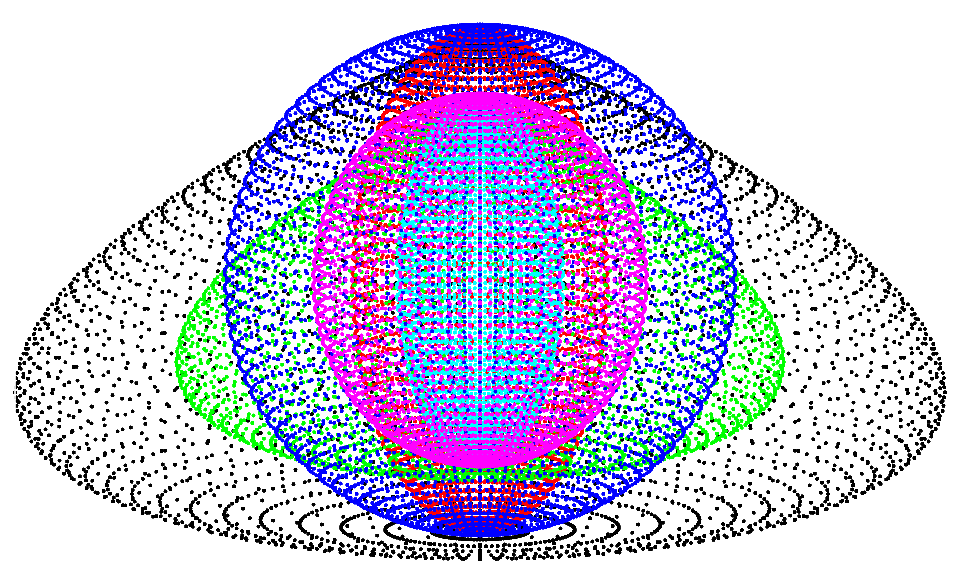}
\end{tabular}
\caption{An object from each class from the 3D synthetic database (scenario 2). \label{figuras_ejemplo21}}
\end{center}
\end{figure}

Fig. (\ref{figuras_ejemplo21}) shows the second scenario, in which there are two "heights" for each class of 3D objects.

The \emph{k-means} algorithm was applied in both scenarios. Once again  the value of parameter $\lambda$ was chosen  as the standard deviation of the points $\{ x_j^k\}_{j=1}^{10000}$, $k=1, \ldots, 30$ (following the criterion also used for the 2D database). The values obtained were $\lambda=0.6237$ in the first scenario and $\lambda=0.7934$ in the second. Once again, the \emph{k-means} algorithm recovered the three groups with the three different geometrical objects in the first scenario, and correctly segmented the figures of the second scenario into six groups, in relation to their "height" and shape.

\section{Application to classifying children's body shapes \label{our_appl}}

The aim of this section is to show how the aforementioned methods can be used to  develop a more efficient apparel sizing
system that can increase accommodation of the population, taking into account the child's body shape. This classification could then be used to choose the most suitable size in a potential online sales application. Before presenting the application, it is fundamental to emphasize a consideration about the  current children sizing system, which is based on the sex and the height or the age of buyers, i.e. when a child wants to buy a T-shirt, he/she has to purchase the size associated with his/her height. It is important to observe that this size is designed for a specific body shape; however, there is a great deal of  variability in body shapes of children with a same height. For example, if the size associated with the child's height does not fit as a consequence of his/her body shape, he/she has to buy the previous or next size of T-shirt, which will probably be too short/long for him/her. In conclusion, it is essential to create a new sizing system which takes into account both: size and shape of the body.\\

A randomly selected sample of Spanish children aged  $3$ to $10$ years old
were
scanned using a Vitus Smart 3D body scanner from Human Solutions. Each child was situated standing and looking forward; and the body shape was  stored as a set of $3075$ $3D$-landmarks (homologous points on his/her surface). The children were asked to wear a standard
white garment in order to standardize the measurements.
From the 3D mesh, several anthropometric measurements were calculated
semi-automatically by combining automatic measurements based
on geometric characteristic points with a manual review.

For each child, the landmarks observed made it possible  to define an oriented smooth triangulated surface immersed in $\mathbb{R}^3$, with a total of $4668$ triangles and the $k$-means algorithm adapted  to the RKHS space can be applied.

The question of the number $k$ of clusters to choose is a difficult problem in data clustering and in particular, in our
application. From the point of view of defining a sizing system, it
is not profitable to design many sizes because it would be very
expensive for apparel companies. On the other hand, because the
objective of an apparel sizing system is to accommodate as large a
percentage of the population as possible, it is not reasonable to
define too few sizes either.

As there is a different size system for each sex, in order to illustrate our procedure the subset of the girls older than 6 was chosen from the whole data set. Children younger than 6 have difficulties in maintaining a standard position during the scanning process, so they were excluded  from our data set. This selection results in a sample of size 195. According to the European standard UNE-EN 13402-3, this age range has  4 different sizes associated with it (1190-1250 mm, 1250-1310 mm, 1310-1370 mm, 1370-1430 mm).

Consequently, we propose two possible sizing system. The first, aims to divide each height range suggested by the UNE-EN 13402-3 standard into two groups (two sizes). We will thus obtain two different sizes for girls with different body shapes within a common height range of. By using this model, most clothes should fit their buyers.  The second sizing system, divides the sample into a lower number of groups (a similar number to that proposed by UNE-EN 13402-3), but bearing in mind the shape and size of the body, not just the height of the children.

Therefore, the body contour from each girl in our data-set is represented by an oriented triangulated smooth surface. For the $j$-th triangle of the surface of the $k$-th girl, its barycenter  $x_{j}^k$ and the normal vector to the surface in $x_j^k$, $\tau_{j}^k$, are calculated. Then, the contour of body of a child is associated with the vector field $\displaystyle \sum_{j=1}^{\textbf{4668}} K(x_j^k, \cdot)(\tau_j^k)$ in $H_K(D, \mathbb{R}^3)$ where $D=[-472.73 , 487.26]\times[-824.72 , 735.28]\times[-156.69, 203.30]$ mm.

Following the same procedure as with the  experimental databases, the \emph{k-means} algorithm with k=2 is applied to each group corresponding to each height range (1190-1250 mm, 1250-1310 mm, 1310-1370 mm, 1370-1430 mm) within the sample. By using $\lambda_1=107.69$, $\lambda_2=178.39$, $\lambda_3=186.98$ and $\lambda_4=196.26$, respectively, in each range, we obtain eight groups, which define the first new recommend sizing system. Each size is described using  the median values (mm) of the anthropometric measurements of the group, as in Table (\ref{tabla_antropometricas_por_alturas}):

\begin{table}[htbp]
\begin{center}
\begin{tabular}{|c|c|c|c|c|c|}
\hline
Size & Height & Chest length & Waist length & Hip length & Group size \\
\hline \hline
 T1 girl & 1209 & 604 & 538 & 656 & 15 \\
 lower (1190-1250)  &  &  &  &  \\ \hline
 T2 girl & 1227 & 670 & 610 & 739 & 29 \\
 upper (1190-1250)  &  &  &  & \\ \hline
 T3 girl & 1273 & 643 & 563 & 688 & 20 \\
 lower (1250-1310)  &  &  &  & \\ \hline
 T4 girl & 1282 & 696.5 & 643.5 & 767 & 31 \\
 upper (1250-1310)  &  &  &  & \\ \hline
 T5 girl & 1331 & 644 & 564.5 & 701 & 32 \\
 lower (1310-1370)  &  &  &  & \\ \hline
 T6 girl & 1346 & 733 & 669 & 807 & 23\\
 upper (1310-1370)  &  &  &  & \\ \hline
 T7 girl & 1393 & 677 & 586 & 750 & 31\\
 lower (1370-1430)  &  &  &  & \\ \hline
 T8 girl & 1410.5 & 797.5 & 722 & 856.5 & 14\\
 upper (1370-1430)  &  &  &  & \\ \hline
\end{tabular}
\caption{First sizing system proposed for heights between 1190 and 1430 mm.}
\label{tabla_antropometricas_por_alturas}
\end{center}
\end{table}

\begin{figure}
\begin{center}
\begin{tabular}{c}
\includegraphics[width=16cm]{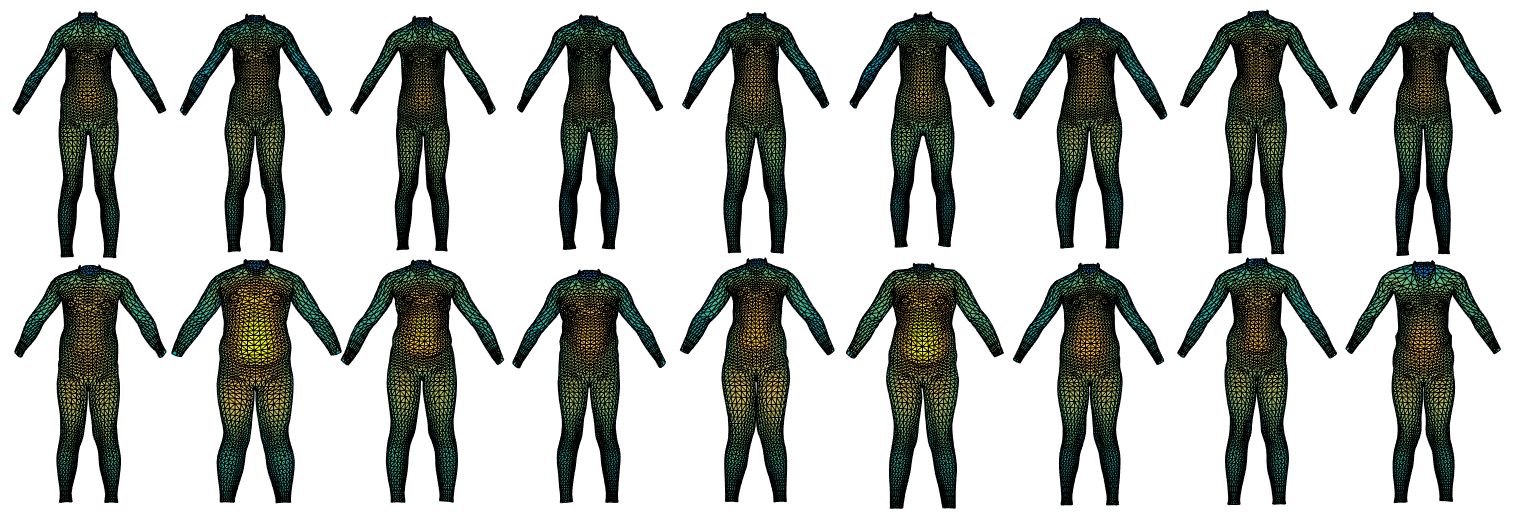}
\end{tabular}
\caption{Girls who are associated with T7 (first row) and T8 (second row) in tthe first sizing system.} \label{altura4}
\end{center}
\end{figure}

Fig. (\ref{altura4}) shows part of a group of girls who belong to the same height range (1370-1430 mm). The bodies of the first row of the image are associated with size T7, and the second row corresponds to T8 in the new sizing system.

\begin{figure}
\begin{center}
\begin{tabular}{cc}
(a)&(b)\\
 \includegraphics[width=6cm]{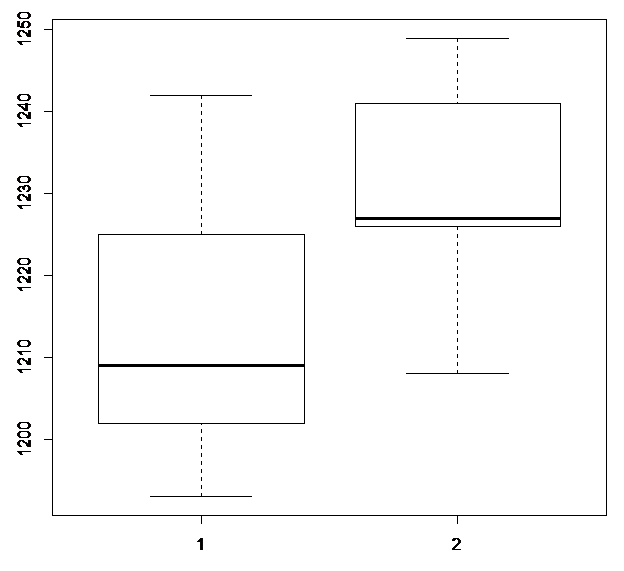} &  \includegraphics[width=6cm]{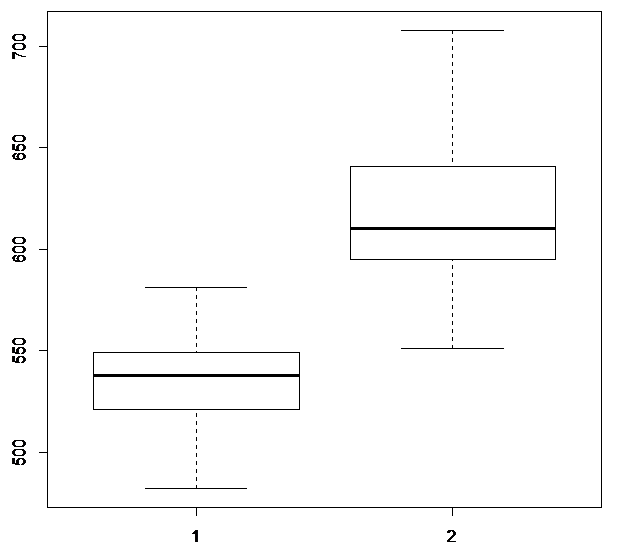} \\
(c)&(d)\\
 \includegraphics[width=6cm]{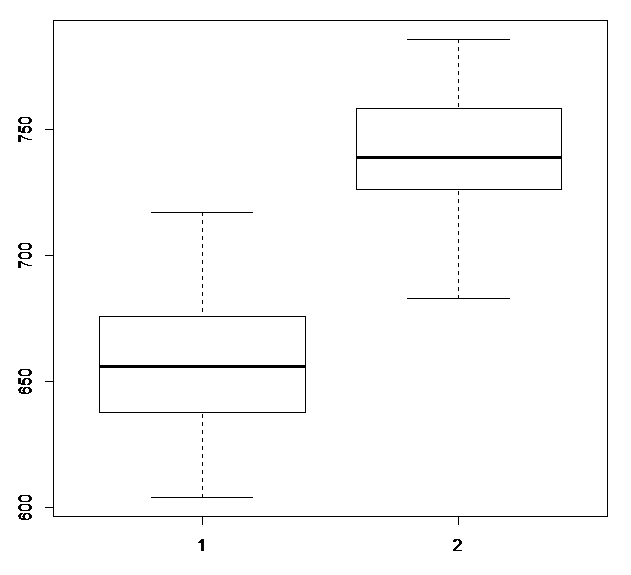} & \includegraphics[width=6cm]{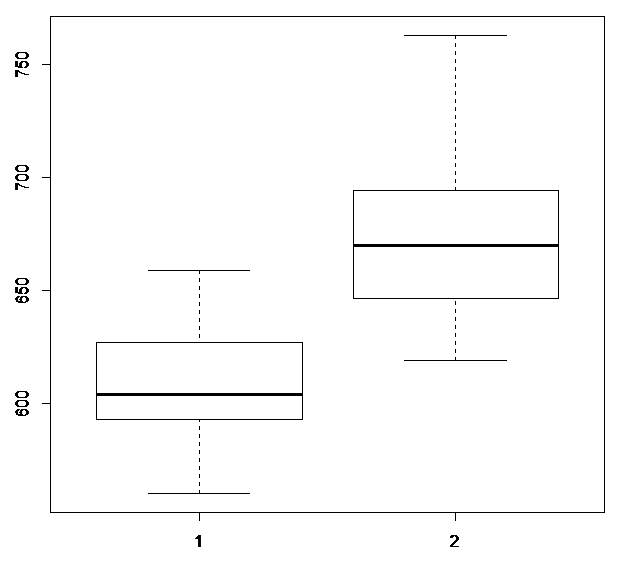}
\end{tabular}
\caption{Height(a), waist length(b), hip length(c) and chest length(d) from the 2 groups obtained (1190-1250mm).}
\label{cajas_por_alturas_1}
\end{center}
\end{figure}

Moreover, in Fig. (\ref{cajas_por_alturas_1}) it is possible to compare some anthropometric measurements of the two groups obtained within the 1190-1250mm height range. Something similar occurs in the height.

To reduce the number of sizes in the previous model, a second model is proposed, which is probably cheaper for the clothing companies. It consists of considering all members of the subsample together and applying \emph{k-means} algorithm to divide the sample into different groups according to the shape and height of the children.

As mentioned above, the question of which number $k$ of clusters to choose is a
difficult problems in data clustering. Several methods have been proposed and used in the literature to make this decision   \cite{Kaufman90,Jain2010}.

In our application we combine the "elbow criterion" with a goodness of clustering measure: the silhouette.

\begin{figure}
\begin{center}
\begin{tabular}{cc}
(a)&(b)\\
 \includegraphics[width=6cm]{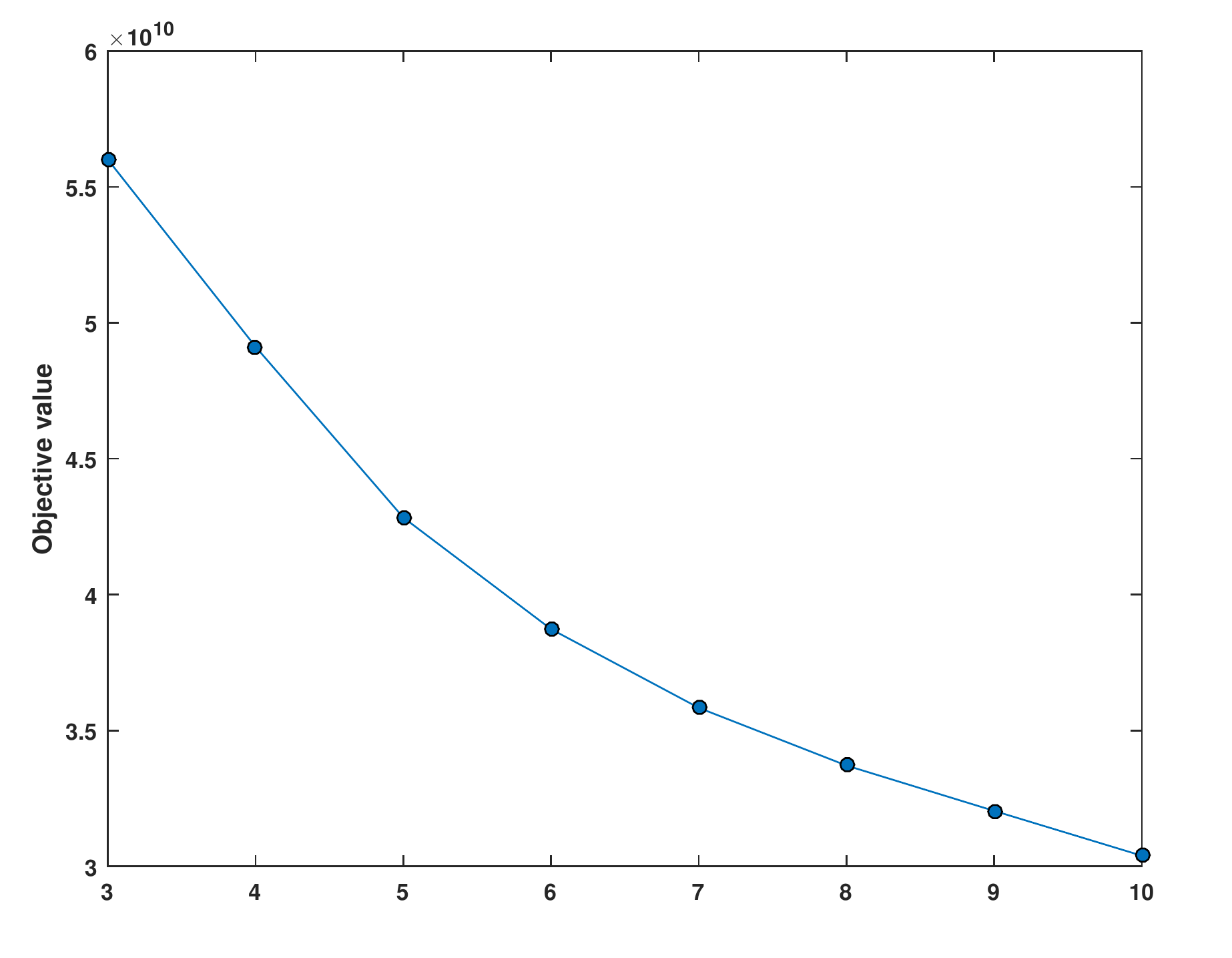} &  \includegraphics[width=6cm]{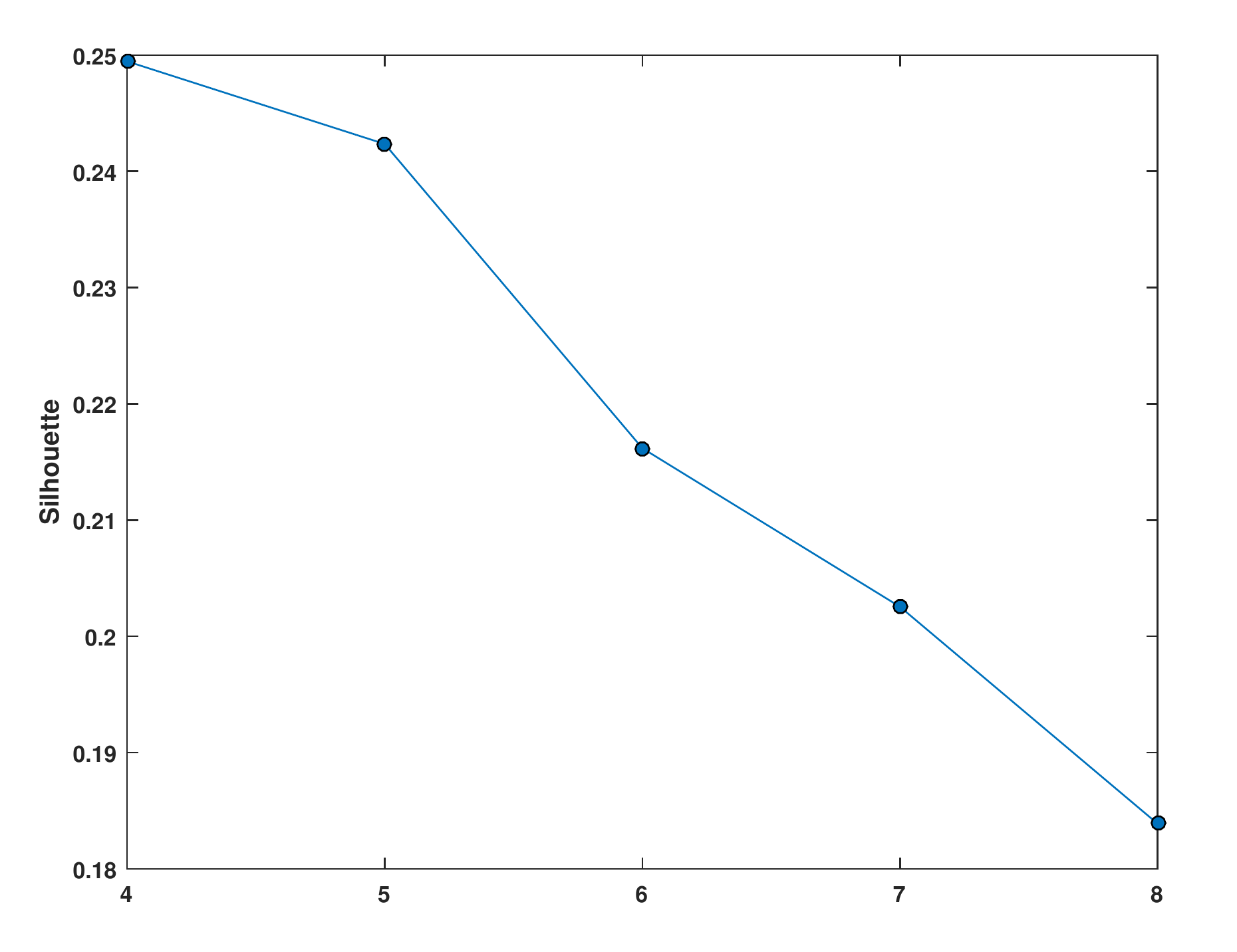}
\end{tabular}
\caption{Study of the "elbow criterion" in (a) and silhouette for different values of k in (b).}
\label{criteriosilueta}
\end{center}
\end{figure}

The idea of the "elbow criterion" is based on
plotting the objective function of Eq. \ref{objetivo1} against the number of clusters. The first clusters greatly decrease the objective function, but at some point the marginal decrement  will drop, giving an angle in the graph.  The number of clusters is chosen at this point, hence the "elbow criterion".

The silhouette of an object is a measure of how close is the object in the neighboring clusters compared with data within its cluster. A silhouette close to 1 implies the datum is in an appropriate cluster, while a silhouette close to -1 implies the datum is in the wrong cluster. The average of the silhouette gives us a measure of the goodness of a clustering.

\begin{figure}
\begin{center}
\begin{tabular}{cc}
(a)&(b)\\
 \includegraphics[width=6cm]{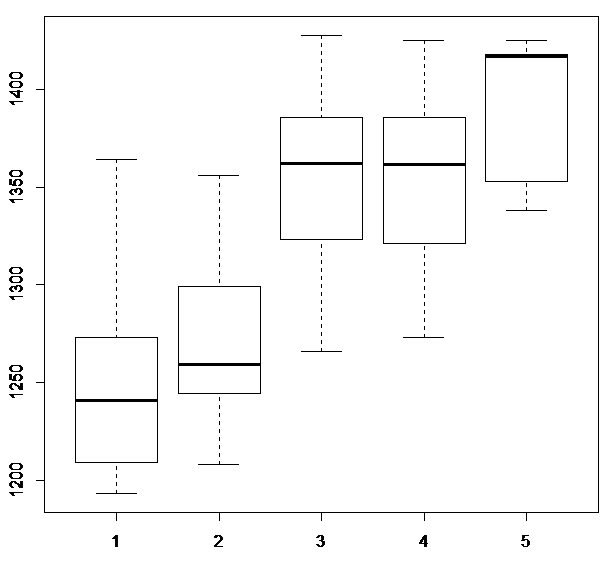} &  \includegraphics[width=6cm]{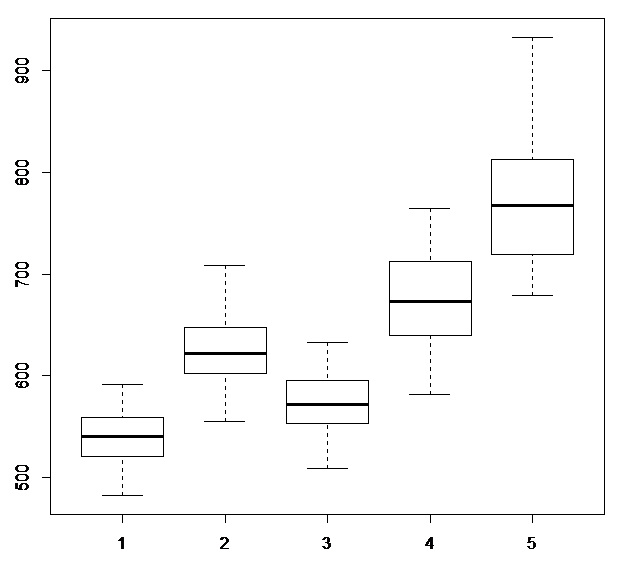} \\
(c)&(d)\\
 \includegraphics[width=6cm]{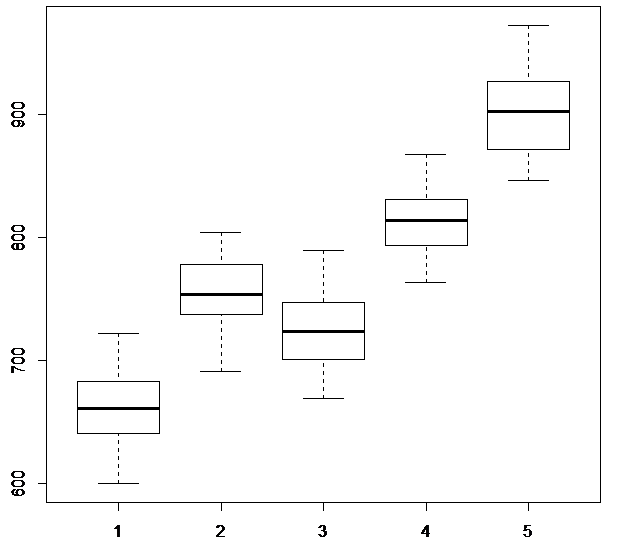} & \includegraphics[width=6cm]{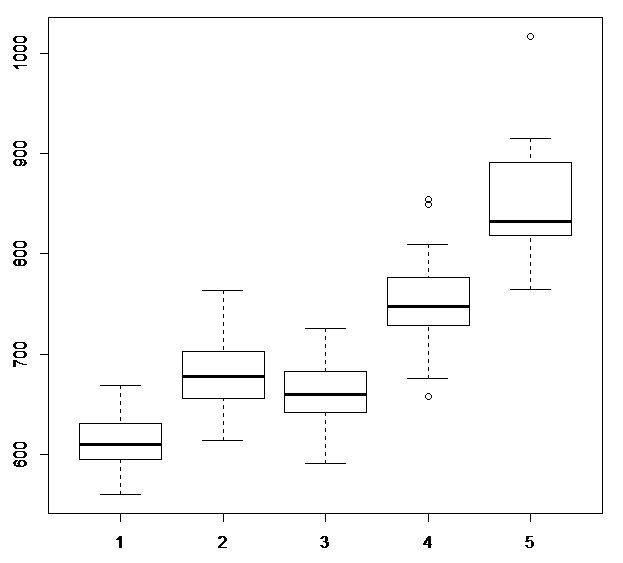}
\end{tabular}
\caption{Height (a), waist length (b), hip length (c) and chest length (d) from the 5 groups obtained (1190-1430mm).}
\label{cajas5}
\end{center}
\end{figure}

In Fig. (\ref{criteriosilueta}) we can see the plot of the objective function (a) and the silhouette (b) against the number of clusters. Taking into account both criteria, we chose to apply the \emph{k-means} algorithm with $k$=5 and $\lambda=183.43$. A partition of the sample is therefore obtained  according to the height and size of the children. Fig.(\ref{cajas5}) shows the box diagrams of some anthropometric measurements within each group generated.

Moreover, we define the new sizing system based on these groups using the median of the anthropometric measurements from each group, as shown in Table (\ref{tabla_antropometricas_alturas_juntas}):

\begin{table}[htbp]
\begin{center}
\begin{tabular}{|c|c|c|c|c|c|}
\hline
Size & Height & Chest length & Waist length & Hip length & Group size \\
\hline \hline
 T1 girl & 1241 & 610 & 540 & 661 & 57 \\ \hline
 T2 girl & 1259 & 678 & 622 & 754 & 39 \\ \hline
 T3 girl & 1362 & 660 & 571.5 & 723.5 & 56 \\ \hline
 T4 girl & 1361.5 & 747.5 & 673 & 814 & 34 \\ \hline
 T5 girl & 1417 & 832 & 767 & 903 & 9 \\ \hline
\end{tabular}
\caption{Second sizing system proposed for heights between 1190 and 1430 mm.}
\label{tabla_antropometricas_alturas_juntas}
\end{center}
\end{table}

\begin{figure}
\begin{center}
\begin{tabular}{c}
\includegraphics[width=16cm]{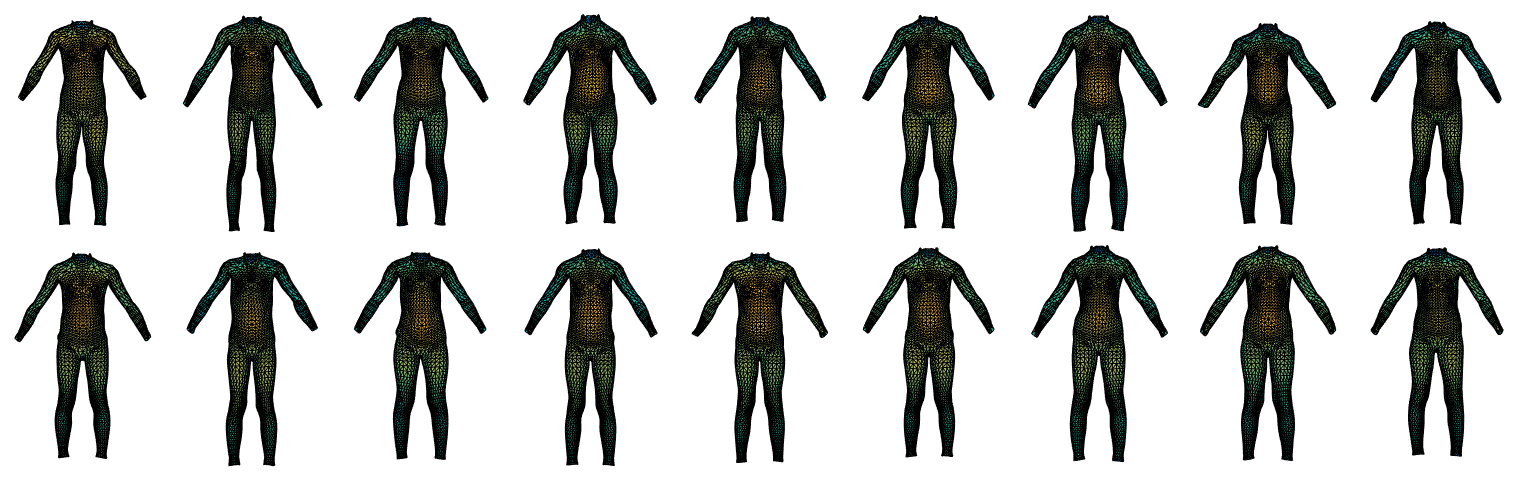}
\end{tabular}
\caption{Girls who are associated with T1 in the second sizing system.} \label{alturajuntasgrupo1}
\end{center}
\end{figure}

Fig. (\ref{alturajuntasgrupo1}) shows part of a group of girls who belong to the same size T1.

\section{Discussion \label{conclusions}}

In this paper we have proposed an approach that represents a novel method
in terms of using the current-based approximation to shape and size analysis in a clustering
procedure and it has been applied in order to define a more efficient children's sizing system. The data are transformed on elements in an RKHS space and the  well-known \emph{k-means}
clustering algorithm has been adapted to it.
An experimental study with simple synthetic objects was successfully conducted to validate the procedure.

We have proposed two ways of defining an efficient sizing system. First, we  segmented the data set
using height,
which is currently the most widely used method. We then applied the
\emph{k-means} algorithm with the number of sizes established  within each
class as $k=2$. In this way, the first segmentation provides a first
easy input to choose the size, while the resulting clusters obtained optimize shape classification within each initial input.
This first classification would provide a large percentage of accommodation but perhaps an excessively large number of sizes. To reduce the number of sizes, in the second system proposed, the \emph{k-means} algorithm is applied to all members of the subsamples and goodness of clustering  criteria were applied to choose the optimal number of groups.
In both cases, the clustering results have been described in terms of  anthropometric dimensions within each group.


\bibliographystyle{plainnat}      
\bibliography{biblio}

\end{document}